\documentclass[aps,pre,preprint,superscriptaddress,showkeys]{revtex4}

\usepackage{graphics}

\begin{document}

\title{Transport optimization on complex networks}

\author{Bogdan Danila}
\email[]{dbogdan@mail.uh.edu}
\affiliation{Department of Physics, The University of Houston, Houston TX 77204-5005}

\author{Yong Yu}
\affiliation{Department of Physics, The University of Houston, Houston TX 77204-5005}

\author{John A.\ Marsh}
\email[]{marshj@ainfosec.com}
\affiliation{Assured Information Security, Rome NY 13440}

\author{Kevin E.\ Bassler}
\email[]{bassler@uh.edu}
\affiliation{Department of Physics, The University of Houston, Houston TX 77204-5005}

\begin{abstract}
We present a comparative study of the application of a recently introduced heuristic algorithm to the optimization of transport on three major types of complex networks. The algorithm balances network traffic iteratively by minimizing the maximum node betweenness with as little path lengthening as possible. We show that by using this optimal routing, a network can sustain significantly higher traffic without jamming than in the case of shortest path routing. A formula is proved that allows quick computation of the average number of hops along the path and of the average travel times once the betweennesses of the nodes are computed. Using this formula, we show that routing optimization preserves the small-world character exhibited by networks under shortest path routing, and that it significantly reduces the average travel time on congested networks with only a negligible increase in the average travel time at low loads. Finally, we study the correlation between the weights of the links in the case of optimal routing and the betweennesses of the nodes connected by them.
\end{abstract}


\maketitle

\noindent {\bf One of the most important problems in the study of complex networks is how to best route transport on the networks. This problem is important because transport is the main function of many natural and human-made networks. Often, the transport routes used on networks are the so-called shortest-path routes, which are the routes with the minimum number of hops between any two nodes. However, this approach, which is currently used to route transport of information packets on the Internet, typically leads to congestion and eventually jamming of highly connected nodes of the networks called hubs. For this reason and in light of recent research, interest has developed in finding the routing rules 
that allow a given network to bear the maximum possible traffic. Specifically, the problem can be stated as follows. Given a complex network and a set of processing power and traffic demand constraints for its nodes, find the set of routing rules which allow the network to bear the highest possible amount of traffic without jamming. This problem is known to be $NP$-hard, meaning that the time required for the computation of an exact solution increases with the number of nodes faster than any polynomial. In this paper we argue that a heuristic transport routing optimization algorithm recently published by us achieves near-optimal transport routing in polynomial time and show this to be true for three important types of complex networks. Of course, any optimized routing when compared to shortest-path routing occurs at the expense of increasing the average number of hops between the nodes. We show that with our algorithm the average number of hops after optimization increases with the number of nodes no faster than logarithmically and that optimization significantly decreases the average travel time on congested networks.
}

\section{Introduction}

\noindent Network transport is a problem encountered in a variety of systems, including biological, social, and a multitude of natural and human-made transport and communication systems. The quantities to be transported can either be of a material nature such as power or goods, or of a non-material nature such as information packets, which are transported on the Internet, or influence, which is transported on social networks \cite{NewmanSIAM, WattsStro,Ericsson,Fortz,GabrelJOH,AllenJOH,MulyanaETT,OurPRE1,YanPRE,Sreenivasan_arXiv, Guimera,Krause,EcheniquePRE,EcheniqueEL,ZhaoLaiParkYe,ParkLaiZhaoYe,TB,TKBHK,OurPRE2, AshtonPRL,Korniss,Anghel,Korniss_arXiv}. Optimization of network transport is thus an important problem for a variety of fields in science and technology. In this paper we present a comparative study of the application of a recently published transport optimization algorithm \cite{OurPRE1} to three major types of complex networks: Erd\H{o}s-R\'enyi \cite{Erdos}, Barab\'asi-Albert \cite{BarabScience}, and uncorrelated scale-free networks generated using the configuration model \cite{Molloy}.

For concreteness, we consider the transport of particles that hop from nodes to nearest-neighbor nodes on complex networks. Traditionally, the routing of network transport is based on the idea of using the shortest paths (usually defined as the paths containing the smallest number of hops) between any two nodes on the network. More generally, the length of a path can be computed as the sum of the weights assigned to the links that form the path. In the case of the Internet for example, link weights are typically assigned manually by operators according to simple rules based on experience \cite{Ericsson}. Recently, a series of algorithms have also been proposed for network traffic optimization. These algorithms are aimed at reducing link \cite{Ericsson,Fortz,GabrelJOH,AllenJOH, MulyanaETT} or node \cite{OurPRE1,YanPRE,Sreenivasan_arXiv,Guimera,Krause} loads by a judicious link weight assignment. They have the effect of improving network transport capacity, which is defined as the rate of particle insertion at which the network becomes jammed.

In a recent paper \cite{OurPRE1} we presented an algorithm that significantly improves transport capacity by a systematic adjustment of link weights to minimize the maximum betweenness on the network. Our algorithm leads to higher transport capacity than other recently proposed algorithms \cite{YanPRE,Sreenivasan_arXiv}. In this paper, we argue that our algorithm achieves near-optimal routing for all three types of complex networks and discuss the reasons why this is possible. Furthermore, we show that routing optimization preserves the small-world character of network routing \cite{WattsStro} and that it significantly decreases the average travel time on congested networks while only marginally increasing it at low loads. Finally, we study the correlation between the optimal weights of the links and the betweenness of the nodes connected by them and show that, as networks approach optimal routing, it becomes impossible to achieve further improvement by relating link weights to node betweennesses.

The problem of finding the exact optimal routing is mathematically tied to the problem of finding the minimal sparsity vertex separator \cite{Sreenivasan_arXiv}, which has been shown to be an $NP$-hard problem \cite{Bui}. This means that the number of flops necessary for the computation of an exact solution increases with the number of nodes $N$ faster than any polynomial. Despite this fact, our heuristic algorithm finds near-optimal solutions for the routing problem in polynomial time. For networks with given average degree, the running time is $O(N^3\log N)$ ($O(N^2\log N)$ for one iteration and requiring $O(N)$ iterations). In its most general form, the algorithm proceeds as follows:

1. Assign uniform or random weight to every link and compute the shortest paths between all pairs of nodes and the betweenness of every node.

2. Find the node which has the highest betweenness $B_{max}$ and increase the weight of every link that connects it to other nodes, or the weight of every incoming link if the network is a directed one. This is done by adding either a constant or a random number to the weight of each link.

3. Recompute the shortest paths and the betweennesses. Go back to step 2.

\noindent Note that the algorithm picks the ``least fit" element of a set and changes its parameters. Therefore, it is a form of extremal optimization \cite{Bak,Boettcher}. However, this algorithm may assign parameters in a deterministic way, unlike many of the other existing extremal optimization algorithms.

The outline of the paper is as follows. In Sec.\ II we give a detailed description of our model and prove a formula that can be used to compute the average number of hops along the path and the average travel time from the betweennesses of the nodes. In Sec.\ III we present our results. Section IV summarizes our results and conclusions.

\section{Model}

We present a comparative analysis of the results obtained with our algorithm \cite{OurPRE1} in the case of three of the most common types of complex networks. These are the random Erd\H{o}s-R\'enyi (ER) networks, Barab\'asi-Albert (BA) networks, and uncorrelated scale-free networks generated using the configuration model (CM). All three network types are undirected. Random ER networks are characterized by a binomial distribution of node degrees, while BA and CM networks exhibit scale-free distributions of node degrees in the limit of a large number of nodes. BA networks are grown by preferential attachment and are characterized by a strong correlation between the degrees of the nodes at the ends of the links. Results are presented for transport for the case of shortest path (SP) routing and for the case of the optimal routing (OR) provided by our algorithm. The number of nodes $N$ varies between 25 and 2500 in the case of SP routing and between 25 and 1600 in the case of optimal routing. To facilitate comparison with previously published results \cite{OurPRE2}, the average degree of both random and BA networks was kept at a constant value of $\left<k\right>=6$, regardless of the number of nodes. Additionally, we consider only fully connected networks. Large connected random networks of lower average degree are prohibitively unlikely to be generated. The average degree of the uncorrelated scale-free networks cannot be strictly controlled but, for a given value of the exponent $\gamma$ that describes the power-law degree distribution of the nodes $p(k)\propto k^{-\gamma}$, it varies with the number of nodes slower than logarithmically. In keeping with Refs.\ \cite{OurPRE1,Sreenivasan_arXiv}, we chose a value of the exponent $\gamma=2.5$, but the value of the lower cutoff parameter \cite{Molloy} was set to $m=4$, which results in the average degree varying between approximately 4.5 and 7.5 as the number of nodes increases from 25 to 1600.

Routing on the network is assumed to be done according to a static protocol which prescribes the next hop(s) for a particle currently at node $i$ and whose destination is node $t$. Each node has a particle queue which works on a ``first-in/first-out" basis. When a new particle is added to the network at some node or arrives at a new node along its path, it is appended at the end of the queue. Upon reaching their destination, particles are removed from the network. For simplicity, we assume that all nodes have the same processing capacity of 1 particle per time step and that new particles are inserted at every node at the same average rate of $r$ particles per time step. This average insertion rate characterizes the load of the network. The destinations of the particles inserted at node $i$ are chosen at random from among the other $N-1$ nodes on the network. The algorithm can, however, be generalized for nodes with different processing capacities and for arbitrary traffic demands.

Given a loop-free routing table, the betweenness $b_i^{(s,t)}$ of node $i$ with respect to a source node $s$ and a destination node $t$ is defined \cite{NewmanPRE} as the sum of the probabilities of all paths between $s$ and $t$ that pass through $i$. The total betweenness $B_i$ is found by summing up the contributions from all pairs of source and destination nodes. The practical way \cite{NewmanPRE} to compute $b_i^{(s,t)}$ for all $i$ and $s$ is as follows: all nodes are assigned weight 1 and then the weight of every node along each path towards $t$ is split evenly among its predecessors in the routing table on the way from $t$ to $s$ and added to the weights of the predecessors. The time average of the number of particles passing through a given node $i$ in the course of a time step is then
\begin{equation}
    \left<w_i\right>_t=\frac{r B_i}{N-1}.
\end{equation}

\noindent Jamming occurs at the critical average insertion rate $r_c$ at which the average number of particles processed by the busiest node reaches unity. Consequently, $r_c$ is given by \cite{Guimera}
\begin{equation}
	r_c=\frac{N-1}{B_{max}},
\end{equation}

\noindent where $B_{max}$ is the highest betweenness of any node on the network. Thus, to achieve optimal routing, the highest betweenness $B_{max}$ should be minimized. An important point is that, even though this minimization procedure pertains to a single scalar quantity, the optimization algorithm implicitly reshapes the betweenness landscape across the whole network, lowering traffic through the initially busy nodes at the expense of increased traffic through the initially idle nodes until the traffic spreads out and an as narrow as possible betweenness distribution is achieved.

To achieve the $O(N^3\log N)$ running time, we used a modified version of the Dijkstra algorithm \cite{Cormen} for the computation of the shortest paths. This version uses binary or Fibonacci heaps to reduce the time required to sort the nodes by distance. We also note that the optimization procedure was started with uniform link weights (SP routing) and half the initial weight was added to the weights of the links connecting the highest betweenness node at every iteration.
\begin{figure}
	\scalebox{0.5}[0.5]{\includegraphics*{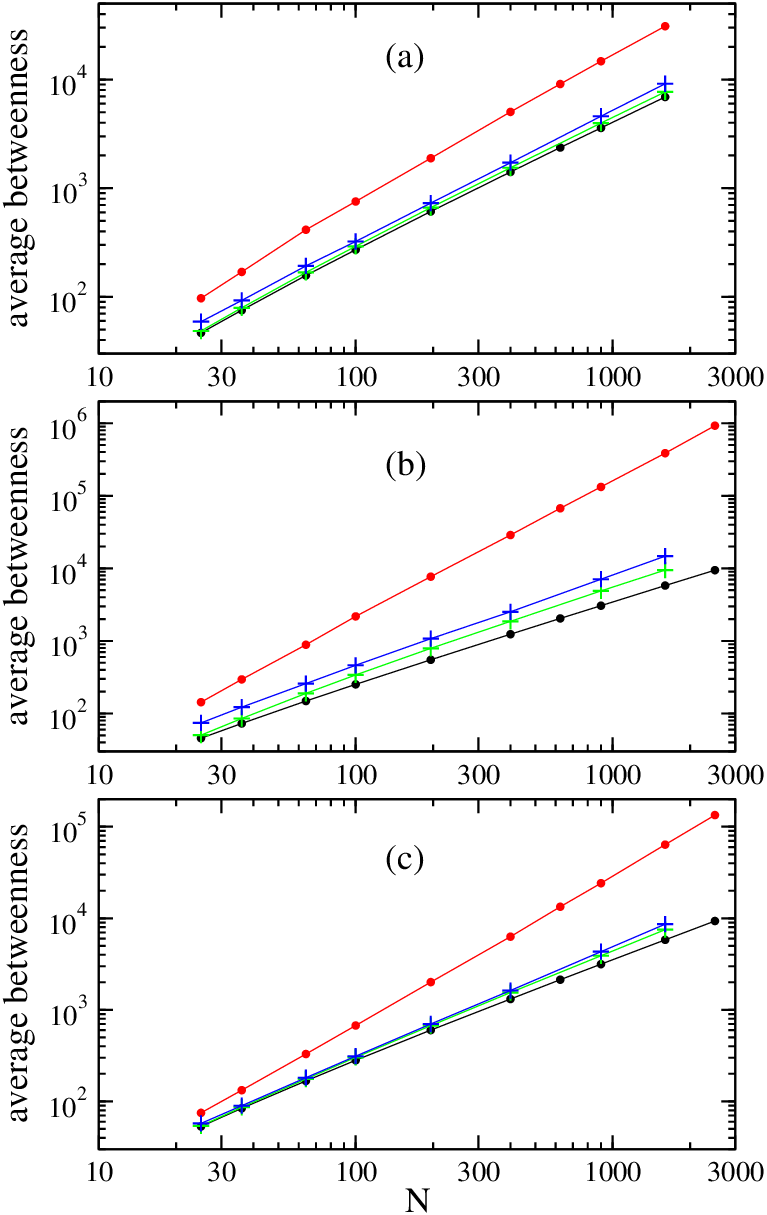}}
	\caption{Ensemble averages of the average and maximum  betweenness as functions of network size for (a) ER, (b) BA, and (c) CM networks. Lower (black) dots represent $\left<B_{avg}^{SP}\right>$, upper (red) dots represent $\left<B_{max}^{SP}\right>$, lower (green) crosses $\left<B_{avg}^{OR}\right>$, and upper (blue) crosses $\left<B_{max}^{OR}\right>$.}
\end{figure}
\begin{figure}
	\scalebox{0.5}[0.5]{\includegraphics*{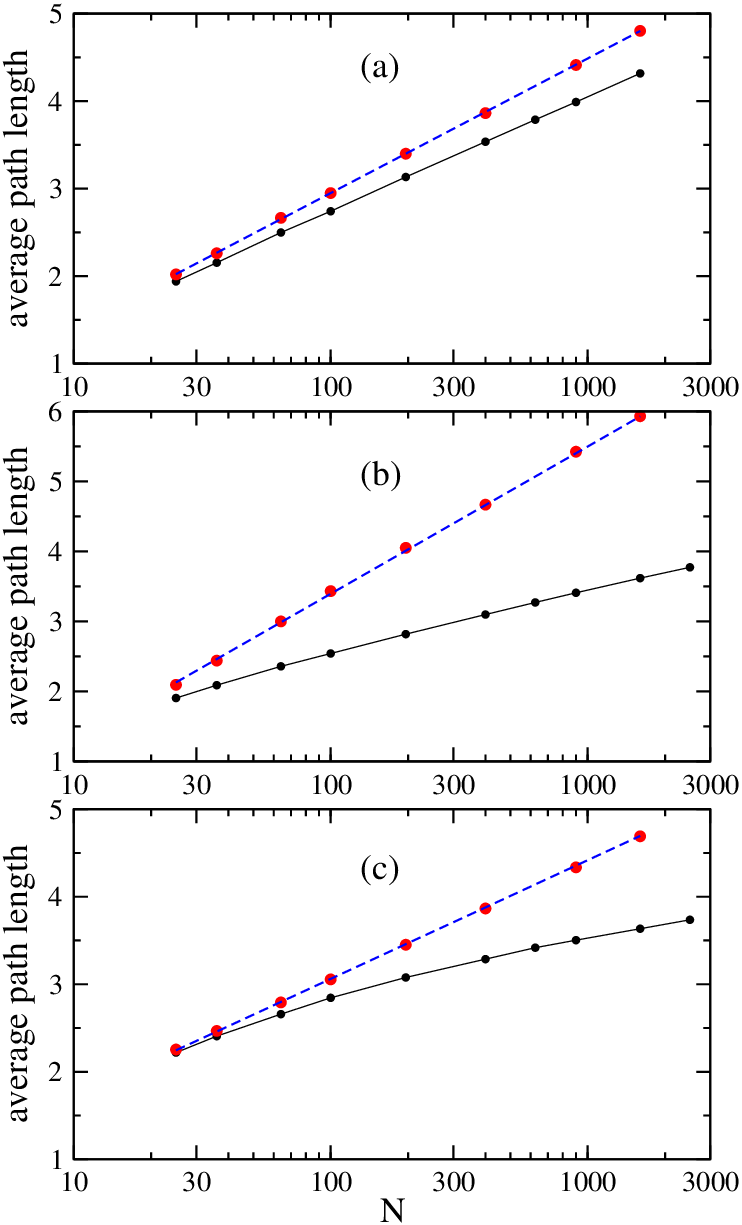}}
	\caption{Ensemble average of the average number of hops along the path as a function of network size for (a) ER, (b) BA, and (c) CM networks. Lower (black) dots represent $\left<L_{avg}^{SP}\right>$, upper (red) squares represent $\left<L_{avg}^{OR}\right>$, and the dashed blue lines represent logarithm fits of the OR points.}
\end{figure}

For the analysis presented in this paper we did not perform the actual transport simulations but instead used analytical formulas to relate the average path length and travel time of the particles to the betweenness values characterizing the various nodes on the network. Before proceeding with the derivation of these formulas we note that, throughout the paper, the network average of a quantity $Q_i$ characterizing the nodes is denoted by $Q_{avg}$, while further averaging over an ensemble of network realizations is indicated by angular brackets.
\begin{figure}
	\scalebox{0.5}[0.5]{\includegraphics*{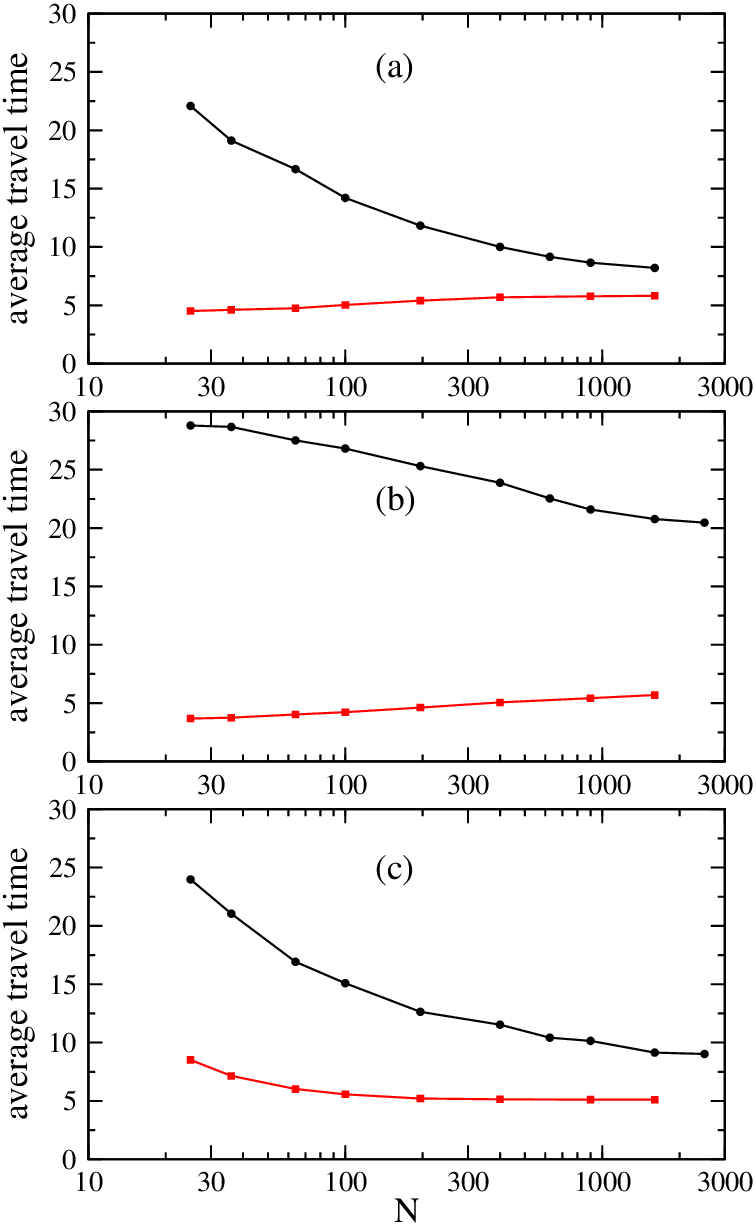}}
	\caption{Ensemble average of the average travel time computed for each network at 99\% of its SP transport capacity, as a function of network size for (a) ER, (b) BA, and (c) CM networks. Upper (black) dots represent $\left<T_{avg}^{SP}\right>$ and lower (red) squares represent $\left<T_{avg}^{OR}\right>$.}
\end{figure}
\begin{figure}
	\scalebox{0.34}[0.34]{\includegraphics*{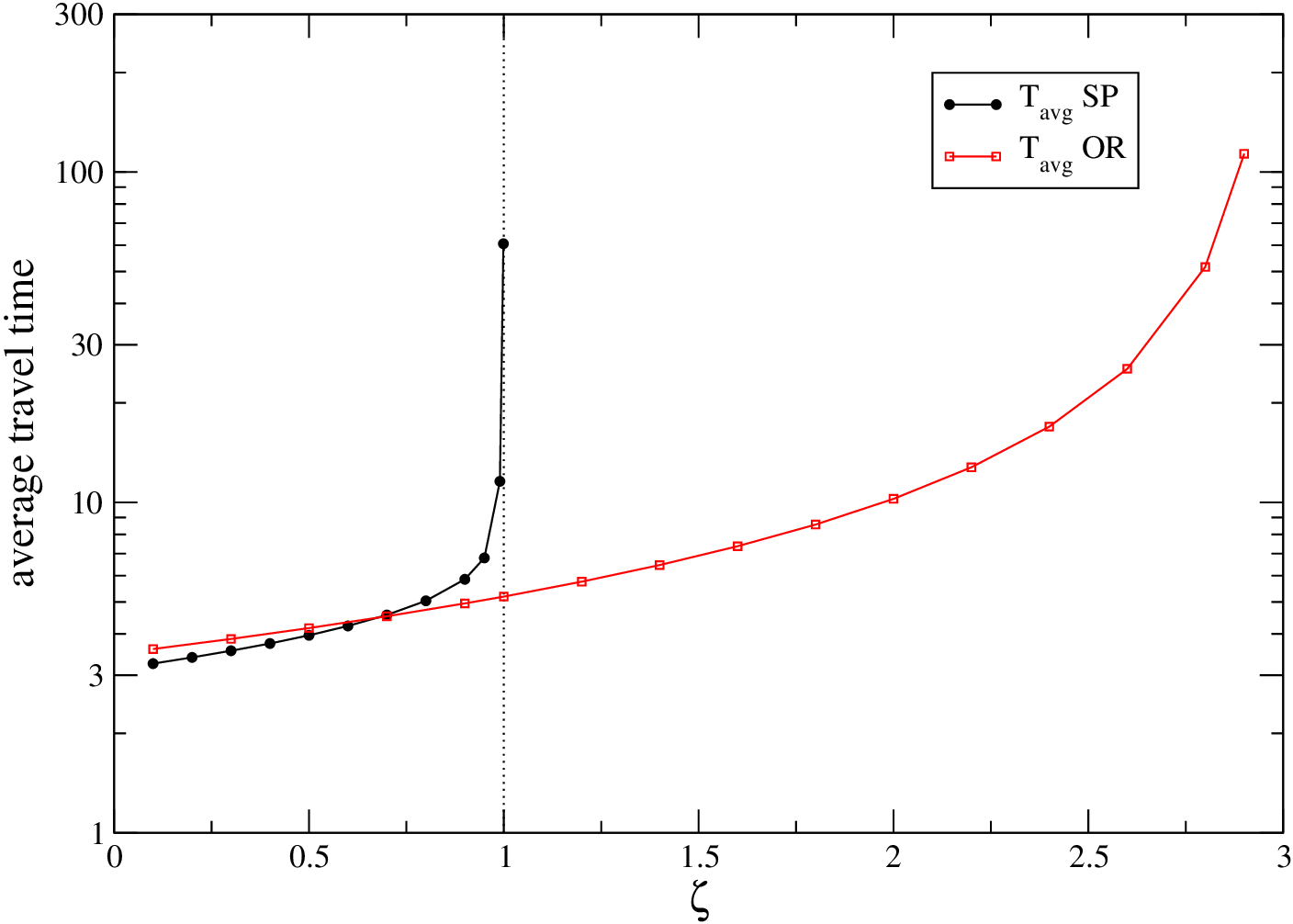}}
	\caption{Average SP (black dots) and OR (red squares) travel times for a CM network with 196 nodes as functions of the load fraction $\zeta$ defined with respect to SP routing.}
\end{figure}

Let $Q_i$ be some quantity associated with each node $i$. Assume we are interested in calculating the average over all paths given by the routing protocol of the sum of $Q_i$ along the path. Let us denote by $p_j(s,t)$ the probability for a particle to be routed along the $j$-th path between $s$ and $t$, and let $\pi_j(s,t)$ be the set of all nodes along that path. In general, the set $\pi_j(s,t)$ may or may not include either $s$ or $t$. The number of possible paths between $s$ and $t$ is $n(s,t)$. Then the betweenness of any node $i$ with respect to $s$ and $t$ is given by

\begin{equation}
    b_i^{(s,t)}=\sum_{j=1}^{n(s,t)} p_j(s,t) \sum_{k\in \pi_j(s,t)} \delta_{ik}.
\end{equation}

Let us now compute the quantity $(\Sigma Q)_{avg}^{(s,t)}$ defined by

\begin{equation}
    (\Sigma Q)_{avg}^{(s,t)}=\sum_{i=1}^N Q_i b_i^{(s,t)}.
\end{equation}

\noindent By substituting Eq.\ (3) into (4) and changing the order of summation, we find

\begin{equation}
    (\Sigma Q)_{avg}^{(s,t)}=\sum_{j=1}^{n(s,t)} p_j(s,t) \sum_{k\in \pi_j(s,t)} Q_k.
\end{equation}

\noindent The inner sum on the right-hand side of Eq.\ (5) is exactly the quantity whose average we are interested in. Thus, Eq.\ (4) gives the average over all possible paths between $s$ and $t$ of the sum of $Q_i$ along the path. Its average over all $s$ and $t$ is $(\Sigma Q)_{avg}$, defined as
\begin{equation}
    (\Sigma Q)_{avg}=\frac{1}{N(N-1)}\sum_{s,t=1}^N (\Sigma Q)_{avg}^{(s,t)}.
\end{equation}

\noindent Using (4), Eq.\ (6) becomes
\begin{equation}
    (\Sigma Q)_{avg}=\frac{1}{N(N-1)}\sum_{i=1}^N Q_i B_i,
\end{equation}

\noindent where $B_i=\sum_{s,t} b_i^{(s,t)}$ is the total betweenness of node $i$. Note that the factor in front of the sum on the right-hand side of Eqs.\ (6) and (7) must be replaced by $1/N^2$ if zero-length paths are counted in the average.

To calculate the average number of hops along the path (which henceforth will be called average path length) $L_{avg}$, we simply set all $Q_i$ equal to 1 and exclude node $t$ from $\pi_j(s,t)$, which is also done in the computation of the betweenness by setting $b_t^{(s,t)}=0$. This leads to
\begin{equation}
    L_{avg}=\frac{1}{N-1} B_{avg}.
\end{equation}

The computation of the average travel time is based on the relationship between the time average of the queue length $\left<q\right>_t$ and the time average of the particle arrival flux $\left<w\right>_t$. It is known from the theory of Markovian queues \cite{Guimera,Allen} that, assuming unity processing power, these quantities are related by
\begin{equation}
    \left<q\right>_t=\frac{\left<w\right>_t}{1-\left<w\right>_t}.
\end{equation}

\noindent In our model, $\left<w\right>_t$ for every node $i$ is given by Eq.\ (1). The average travel times are computed at a fraction $\zeta$ of the the critical insertion rate $r_c^{SP}=(N-1)/B_{max}^{SP}$ at which the network starts jamming when using shortest path routing. Thus, we have
\begin{equation}
    \left<w_i\right>_t=\zeta\frac{B_i}{B_{max}^{SP}},
\end{equation}

\noindent which yields
\begin{equation}
    \left<q_i\right>_t=\frac{\zeta B_i}{B_{max}^{SP}-\zeta B_i}.
\end{equation}

\noindent Unless otherwise specified, all average travel times were computed for $\zeta=0.99$.

The quantity associated with every node $i$ is in this case $Q_i=T_i=1+\left<q_i\right>_t$. This accounts for the average number of time steps a particle has to wait in the queue of node $i$ plus one time step to hop to the next node. When the resulting expression for $Q_i$ is substituted into Eq.\ (7), we find
\begin{equation}
    T_{avg}=\frac{B_{max}^{SP}}{N(N-1)} \sum_{i=1}^N \frac{B_i}{B_{max}^{SP}-\zeta B_i}.
\end{equation}

\section{Results}

Fig.\ 1 shows plots of the SP and OR ensemble averages of the network average and maximum betweenness, $\left<B_{avg}\right>$ and $\left<B_{max}\right>$ respectively, as functions of network size $N$. All ensemble averages are computed over a set of 100 network realizations. Fig.\ 1(a) shows the results for random networks, while Figs.\ 1(b,c) pertain to BA and CM networks, respectively. In light of Eq.\ (8) and of the fact that the average path lengths of all three types of networks are known to increase with network size no faster than $\log N$, we expect their average SP betweenness to increase no faster than $N\log N$. The maximum SP betweenness is known to scale with network size according to a power law \cite{OurPRE1,Sreenivasan_arXiv}. Our results show that, regardless of network type, the same types of laws characterize the average and maximum betweenness after optimization. Results for the exponents of the power laws characterizing $\left<B_{max}\right>$ for the six network type and routing combinations are given in Table I, with the quoted errors being 2$\sigma$ estimates. The exponents for $\left<B_{max}\right>$ in the case of random networks were obtained by fitting data corresponding to $N$ between 64 and 1600, while all other exponents were obtained for $N$ between 25 and 1600.
\begin{table}
\begin{center}
	\begin{tabular}{|c||c||c|}
	\hline
		  & SP & OR \\
	\hline
    ER & $1.381\pm 0.017$ & $1.214\pm 0.022$ \\
    BA & $1.897\pm 0.008$ & $1.273\pm 0.009$ \\
    CM & $1.626\pm 0.011$ & $1.207\pm 0.009$ \\
	\hline
	\end{tabular}
	\caption{Exponents of the $\left<B_{max}\right>$ power-law scaling with network size $N$ for Erd\H{o}s-R\'enyi, Barab\'asi-Albert and configuration model networks, before (SP) and after (OR) optimization.}
\end{center}
\end{table}

\begin{figure}
	\scalebox{0.5}[0.5]{\includegraphics*{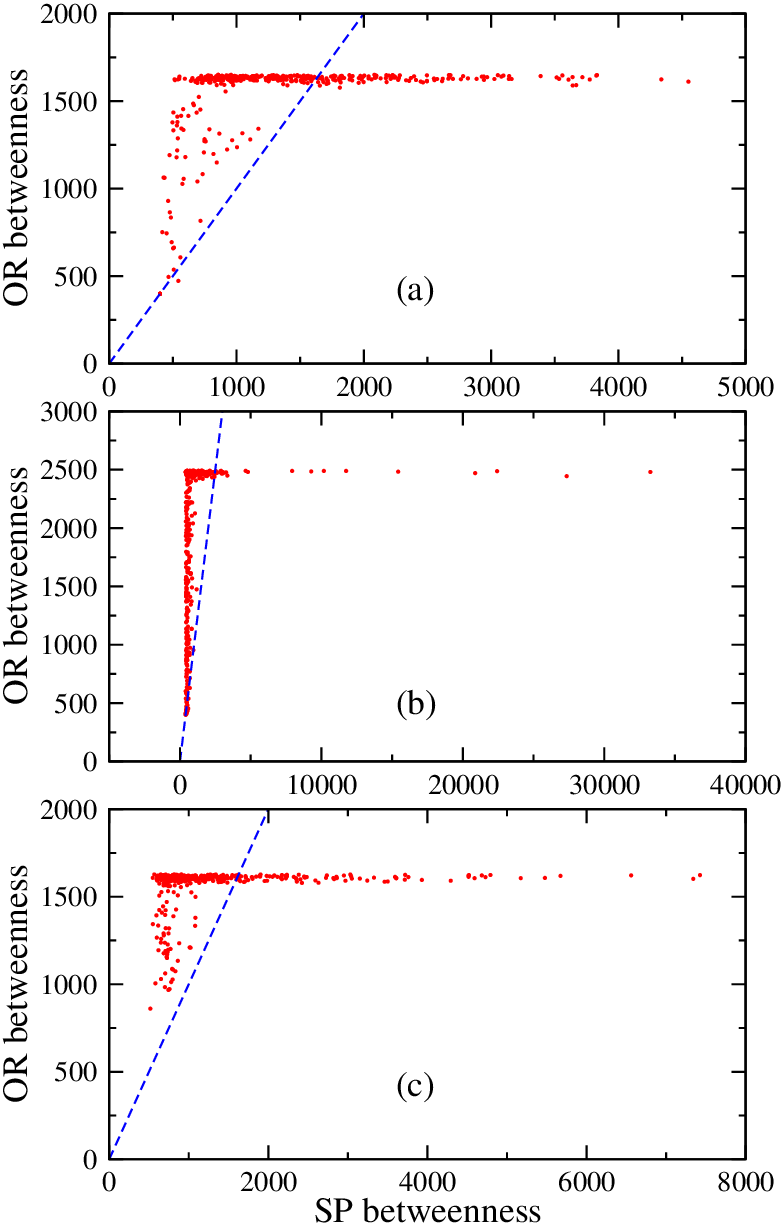}}
	\caption{Correlation plots of the final (OR) versus initial (SP) betweenness for networks with 400 nodes. Results are for (a) an ER network, (b) a BA network, and (c) a CM network.}
\end{figure}

It is apparent from Fig.\ 1 that our optimization algorithm lowers the exponents of the power laws significantly, leading to far smaller values of the maximum betweenness than in the case of SP routing. As a result, the transport capacity, quantified by the critical insertion rate $r_c$, is much higher. The lower values of the OR exponents in Table I also mean that transport capacity for networks with nodes of given processing power decreases much slower with increasing network size. Moreover, even though the ensemble average of the maximum betweenness scales with $N$ according to a power law while the ensemble average of the average betweenness is proportional to $N \log N$, the difference between their OR values remains negligible over almost two orders of magnitude of network size. This indicates the optimality of the routing. Finally, optimization leads to only a small increase in the average betweenness (which is explained by the need to have slightly longer paths around the hubs). The reason for the higher values of the exponents exhibited by Barab\'asi-Albert networks both before and after optimization is discussed in a later paragraph.

Plots of the ensemble average of the average path length $\left<L_{avg}\right>$ as a function of network size are shown in Fig.\ 2. As expected, the average SP path length is proportional to $\log N$ in the case of ER networks and increases even slower with network size in the case of the scale-free networks. The important finding is that after optimization, the average path length scales with the logarithm of network size for all three types of networks. This means that routing optimization preserves the small-world character of network routing \cite{WattsStro}.
\begin{figure*}
	\scalebox{0.6}[0.6]{\includegraphics*{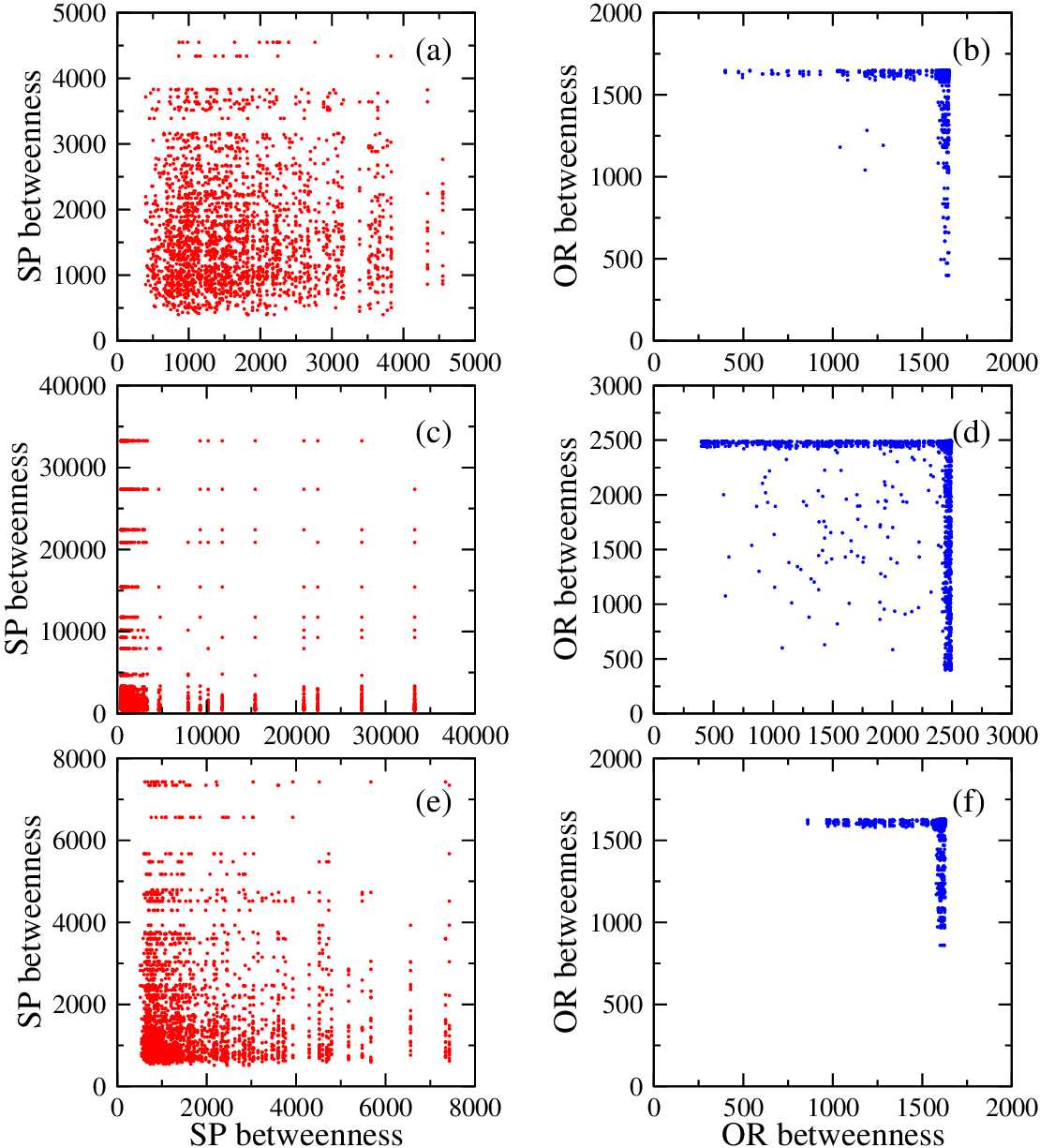}}
	\caption{Correlation plots of the betweenness of each node with the betweenness of its neighbors in the case of SP routing [(a), (c), and (e)] and after optimization [(b), (d), and (f)]. Results are for an ER network [(a), (b)], a BA network [(c), (d)], and a CM network [(e), (f)], each with 400 nodes.}
\end{figure*}

In Fig.\ 3 we show the network size dependence of the ensemble average of the average travel time $\left<T_{avg}\right>$. Average travel times are computed for each network realization using Eq.\ (12) at 99\% of the critical insertion rate corresponding to shortest path routing. It is apparent that, regardless of network type or size, when routing optimization is applied to a network working close to its maximum transport capacity, it results in significant reduction of the average travel time between source and destination. This is in addition to the fact that optimization allows insertion rates significantly higher than the critical rate for SP routing.

The dependence of the average travel time on network load is shown in Fig.\ 4, where $T_{avg}$ for a single CM network realization with 196 nodes is plotted against the load parameter $\zeta=r/r_c^{SP}$. For this case, the ratio between the SP and OR maximum betweennesses is approximately 2.95, which is the maximum allowable value for $\zeta$ when this network uses optimal routing. Even though optimal routing results in longer travel times when the network bears only small loads, the increase is not significant. For the overall efficiency of network transport, this is at least as important as the decrease in travel time at higher loads.

Plots of the optimal routing (OR) betweenness versus the shortest path (SP) betweenness for one network with $N$=400 nodes of each type are shown in Fig.\ 5. It is apparent from these plots that the algorithm performs remarkably well, lowering traffic through all nodes whose initial betweenness lies above a certain critical value until they all reach essentially the same ``critical" betweenness. On the other hand, virtually all nodes whose SP betweenness lies below the critical value experience higher traffic, many of them (especially those with higher initial betweenness) reaching the critical value. Therefore it is unlikely that another optimization algorithm could achieve a significantly lower critical betweenness by further diverting traffic towards some of the nodes which still have below-critical betweenness. This is because not all low betweenness nodes can have their betweenness increased without unduly lengthening paths or increasing traffic through other nodes which are prone to congestion. The simplest examples (which are not valid in the case of BA networks) are those of a small subnetwork connected to the rest of the network through a single link to a high SP betweenness node, or a triangle connected to the rest of the network only by containing such a node. There is no way of diverting traffic through the aforementioned structures, and nodes belonging to them will have low betweenness even in the case of rigorously optimal routing.
\begin{figure*}
	\scalebox{0.6}[0.6]{\includegraphics*{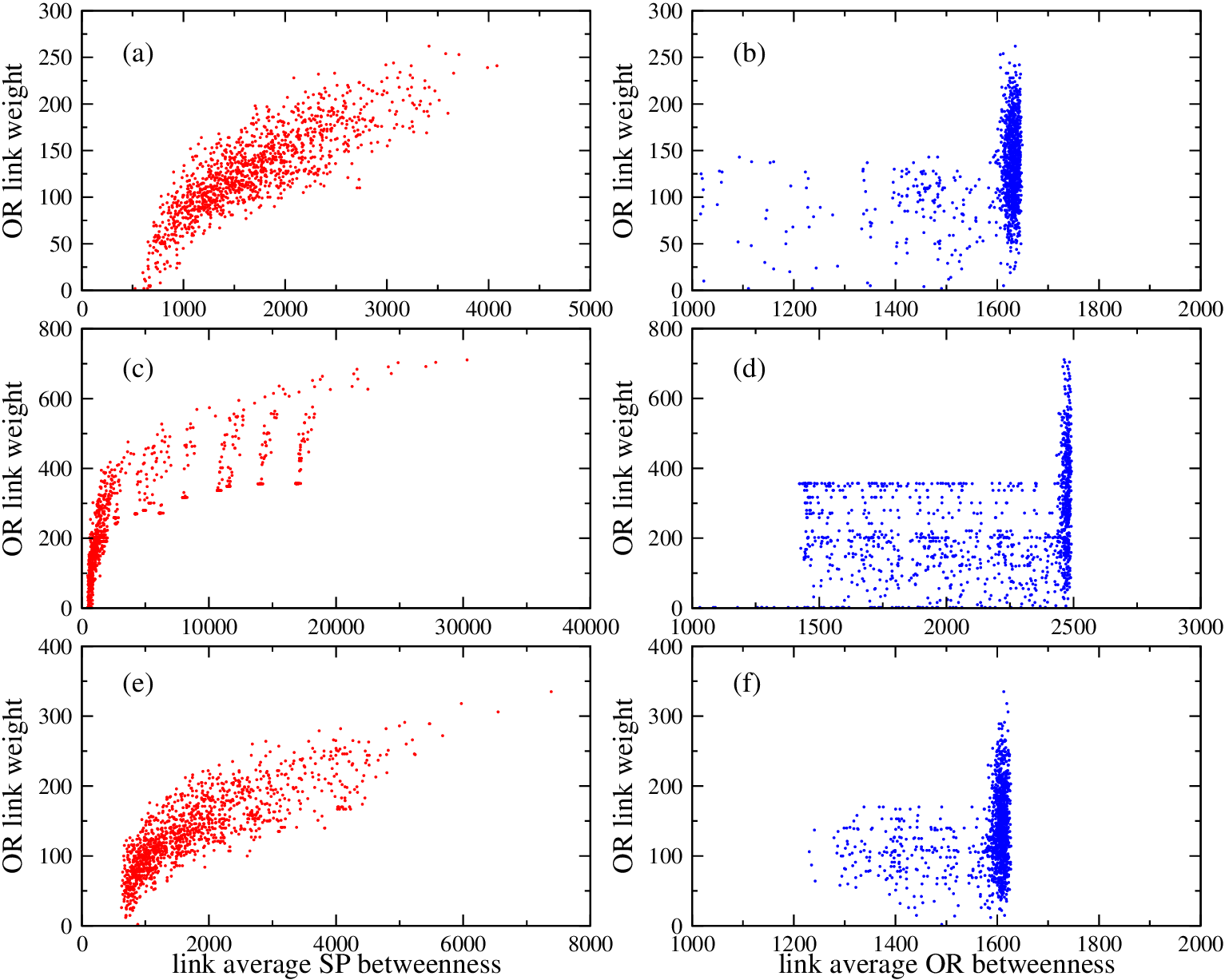}}
	\caption{Correlation plots of the final (OR) link weights versus link average SP betweenness [(a), (c), and (e)] and link average OR betweenness [(b), (d), and (f)]. Results are for an ER network [(a), (b)], a BA network [(c), (d)], and a CM network [(e), (f)], each with 400 nodes.}
\end{figure*}

The case of Barab\'asi-Albert networks deserves special attention. As can be seen from Table I, their power-law exponent is higher both before and after optimization. Moreover, the initial betweenness is spread over a much wider interval, and even after optimization there is a narrow but dense ``trail" of nodes of the lowest possible SP betweenness and whose OR betweenness is still far from the critical value. This behavior is explained by the peculiar structure of BA networks, which is due to the way they are grown. As shown in \cite{OurPRE2}, they consist basically of three categories of nodes. The first category comprises nodes of high degree which are likely to be connected to each other. These are mainly the nodes which formed the ``core" of the network and some nodes that were attached to them in the early stages of network growth. The second category comprises the multitude of low degree ``latecomers" which did not have the chance for another node to be attached to them in the process of growing the network by preferential attachment. These nodes are connected only to nodes in the first category but not to each other. Their SP betweenness is generally at or very close to the lowest possible value. On the other hand, traffic between them invariably passes through some of the nodes from the first category, further increasing the betweenness of the latter. Finally, there is a third category of nodes of intermediate degree which are connected mainly to those in the first category but also sparsely connected to each other. Their connections, however, are not sufficiently many to form large connected subnetworks. Their betweenness is distributed over a range which is narrower than in the case of the nodes on a random network of the same size and average degree. The ratio between the number of nodes in the third and second category increases with network size. At first, the optimization algorithm is successful in diverting traffic away from the highest betweenness nodes by using other nodes from the first category. However, when it tries to find alternative paths between high betweennesses nodes through nodes from the second or third category, it runs into problems. All too often, such an alternative path must go back and forth between the first and the other two sets of nodes, thus being likely to contain a high betweenness node to be avoided. For this reason, nodes from the second or third category are unlikely to be useful as part of alternative paths and their betweenness remains relatively low after optimization. These considerations explain the higher SP and OR maximum betweennesses of BA networks as well as the dense trail of nodes at the lowest SP betweenness in Fig.\ 5(b). Thus, from the point of view of both shortest path as well as optimal routing, BA networks are by far the worst. If a scale-free topology is desired or unavoidable \cite{TB}, the network should be structured as close as possible to an uncorrelated one. An interesting question is whether biological or social transport networks exhibit any correlation between the degrees of the nodes connected by links, and whether evolutionary mechanisms are capable of avoiding such correlations for the sake of improved transport capacity.

The above considerations are illustrated in Fig.\ 6, where we present correlation plots of the betweenness of each node with the betweennesses of its neighbors. The correlations of the SP betweennesses are shown in Figs.\ 6(a,c,e) while the correlations of the OR betweennesses are shown in Figs.\ 6(b,d,f). In the case of random networks (which are by definition uncorrelated) as well as in the case of uncorrelated scale-free networks, the SP-SP correlation plots are consistent with a probability density of the dots (representing links) proportional to the product of the probabilities of having end nodes of given betweennesses. On the other hand, the SP-SP correlation plot for BA networks exhibits areas of high density near the axes corresponding to links between a node from the first category and another one from the second or third, a small but relatively dense patch next to the origin corresponding to links within the third category, and a low density but relatively uniform distribution of links between high betweenness nodes. After optimization, the two uncorrelated networks (Figs.\ 6 (b) and (f)) exhibit only links between two nodes close to the critical betweenness or between a node close to critical betweenness and a lower betweenness node, and the density of the low betweenness nodes decreases quickly when moving away from the critical betweenness. On the other hand, links between two nodes from the third category mentioned above, whose betweennesses remain well below critical, can be seen in the case of the BA network in Fig.\ 6(d). Moreover, the density of the links between a near-critical betweenness node and a lower betweenness one is independent of the lower betweenness value.

Intuitively, one may expect the final (OR) link weights to be simply related to the average of the initial (SP) betweenness values of the two adjoined nodes. (Or, if considering directed networks, it would make sense to use the betweenness of the destination node.) To study this possibility, we have examined the correlation between the OR link weights and the average betweennesses values of the two adjoined nodes. Results for correlations against both the average SP betweenness and the average OR betweenness are presented in Fig.\ 7. It is apparent from Figs.\ 7(a,c,e) that the correlation between OR link weight and average SP betweenness, while notable, is neither strong, nor linear. The correlation is particularly weak in the case of Barab\'asi-Albert networks. This explains why faster, non-iterative optimization algorithms like the one described in Ref.\ \cite{YanPRE}, which assume a direct proportionality between link weight and node degree or shortest path betweenness, do work to some extent but still leave a lot of room for improvement. It is worth noting that the correlation between link weight and betweenness does not improve if one uses the maximum of the betweennesses of the two nodes connected by the link, nor when using an undirected version of our routing algorithm, which increases only the weights of the links incoming to the highest betweenness node at every iteration. Moreover, a previous paper \cite{Krause} presenting an iterative routing optimization algorithm which updates betweenness globally at every iteration by setting link weights proportional to the betweenness of the destination node has reported no improvement after the first three to four iterations. Subsequent iterations do shuffle traffic around, but without further reducing the maximum betweenness. This is explained by Figs.\ 7(b,d,f) which show that any correlation between link weight and betweenness is destroyed as the network approaches optimal routing. Consequently, once routing is sufficiently close to optimal, it becomes impossible to achieve further incremental improvements by relating link weights to node betweennesses.

\section{Conclusions}

In summary, we have presented a simple heuristic algorithm for routing optimization on networks and demonstrated its usefulness for three major types of complex networks. Results show that the application of this algorithm allows all three types of networks to bear significantly higher traffic than in the case of shortest path routing. Network transport capacity is improved by a factor which increases with network size according to a power law. The best results are obtained in the case of uncorrelated networks, especially those with a scale-free distribution of node degrees. The Barab\'asi-Albert algorithm of growth by preferential attachment leads to networks which are extremely prone to congestion when using a shortest path routing protocol and, while our routing optimization algorithm is able to correct the problem to a great extent, such networks are still at a disadvantage after optimization. The explanation of this fact resides in the highly correlated fashion in which links are assigned when growing BA networks.

We have found a simple analytical formula (7) which allows the calculation of the average of the sum along the path of any quantity characterizing the nodes. In particular, this formula can be used to compute average path lengths and travel times. We have shown that the unavoidable lengthening of the paths due to routing optimization still preserves the small-world character of the network exhibited in the case of shortest-path routing. More important, optimal routing leads to much shorter average travel times than its shortest path counterpart at load levels at which a network using SP routing becomes congested, while the lengthening of the average travel times at low loads is negligible.

Finally, we show that there is no correlation between the optimal weight of a link and the optimal routing betweenness of the nodes at its ends, and that the correlation is weak and nonlinear if shortest path betweenness is used. This explains the performance limitations of previously proposed routing optimization algorithms, which attempt to determine link weights from node betweennesses. The only way to avoid this limitation is to update link weights incrementally, and only for the links connecting the node which exhibits the highest betweenness at the previous iteration.

\begin{acknowledgments}

The authors acknowledge support from the NSF through grant No.\ DMR-0427538 and also from SI International through A.\ Williams of the Air Force Research Laboratory Information Directorate under contract No.\ FA8750-04-C-0258.

\end{acknowledgments}

\end{document}